\documentclass[aps,showpacs,preprintnumbers,amsmath, amssymb]{revtex4}

\usepackage{float}
\usepackage{graphics,epsfig}
\usepackage{graphicx}
\usepackage{dcolumn}
\usepackage{bm}
\usepackage{epstopdf}
\begin{document}
\baselineskip=0.8 cm
\title{{\bf Holographic superfluid with excited states}}

\author{Dong Wang$^{1}$, Qiyuan Pan$^{1,2,3}$\footnote{panqiyuan@hunnu.edu.cn}, Chuyu Lai$^{4}$\footnote{laichuyu@gzhu.edu.cn}, and Jiliang Jing$^{1,2,3}$\footnote{jljing@hunnu.edu.cn}}
\affiliation{$^{1}$ Key Laboratory of Low Dimensional Quantum Structures and Quantum Control of Ministry of Education, Synergetic Innovation Center for Quantum Effects and Applications, and Department of Physics, Hunan Normal University, Changsha, Hunan
410081, China}
\affiliation{$^{2}$ Institute of Interdisciplinary Studies, Hunan Normal University, Changsha, Hunan 410081, China}
\affiliation{$^{3}$ Center for Gravitation and Cosmology, College of Physical Science and Technology, Yangzhou University, Yangzhou 225009, China}
\affiliation{$^{4}$ Center for Astrophysics, School of Physics and Materials Science, Guangzhou University, Guangzhou 510006, China}

\vspace*{0.2cm}
\begin{abstract}
\baselineskip=0.6 cm
\begin{center}
{\bf Abstract}
\end{center}

We construct a novel family of solutions of the holographic superfluid model with the excited states in the probe limit. We observe that the higher excited state or larger superfluid velocity will make the scalar hair more difficult to be developed, and the higher excited state or smaller mass of the scalar field makes it easier for the emergence of translating point from the second-order transition to the first-order one. We note that the difference of the critical chemical potential between the consecutive states increases as the superfluid velocity increases. Interestingly, the ``Cave of Winds" phase structure will disappear but the first-order phase transition occurs for the excited states, which is completely different from the holographic superfluid model with the ground state. This means that the excited state will hinder the appearance of the Cave of Winds. Moreover, we find that there exist additional poles in Im[$\sigma(\omega)$] and delta functions in Re[$\sigma(\omega)$] arising at low temperature for the excited states, and the higher excited state or larger superfluid velocity results in the larger deviation from the expected relation in the gap frequency.

\end{abstract}

%\keywords{AdS/CFT correspondence, Holography and condensed matter physics (AdS/CMT), Holographic superfluid, Excited states}

\maketitle
\newpage
\vspace*{0.2cm}
\section{Introduction}

Nowadays, the phenomena of high-temperature superconductivity lack a widely-accepted microscopic explanation by conventional approaches due to the strong coupling \cite{Hartnoll2009,Herzog2009,Horowitz2011}. Since the gauge/gravity duality relates
the strong coupling conformal field theory in $d$-dimensions to a weak gravitational system in ($d+1$)-dimensions \cite{Maldacena,Gubser1998,Witten}, it is widely believed that this holographic method can provide novel and effective ways to investigate universal properties of high-temperature superconductors \cite{CaiQongGen2015}. Considering the spontaneous symmetry breaking of the $U(1)$ symmetry of an Abelian Higgs model coupled to the anti-de Sitter (AdS) gravity \cite{GubserPRD78}, Hartnoll \emph{et al.} built the so-called holographic $s$-wave superconductor model and used this simple ($3+1$)-dimensional bulk theory to reproduce properties of a ($2+1$)-dimensional superconductor \cite{HartnollPRL}. Taking the backreaction of matter fields on the spacetime metric into account, they found that the backreaction makes condensation harder and even the uncharged scalar field can form a condensate \cite{HartnollJHEP12}. To go a step further, by turning on the spatial component of the gauge field, Basu \emph{et al.} constructed a supercurrent solution in AdS$_{4}$ and obtained a ``special point" (critical point) in the phase diagram where the second order superconducting phase transition will become first order \cite{BasuMukherjeeShieh}. In Ref. \cite{HerzogKovtunSon}, this interesting superfluid phase was also discussed by Herzog \emph{et al.} in detail. Generalizing Refs. \cite{BasuMukherjeeShieh,HerzogKovtunSon} to various scalar masses and also considering AdS$_{5}$ as well, Arean \emph{et al.} observed the peculiar shape of the corresponding condensate curve, i.e., the system first suffers a second-order transition and then a first-order phase transition when the temperature decreases, and called it the ``Cave of Winds" \cite{AreanJHEP}. Interestingly, for the $p$-wave superfluid via the Maxwell complex vector field model \cite{CaiPWave-1,CaiPWave-2}, the Cave of Winds phase structure also appears \cite{PWavSuperfluid,LifshitzSuperfluid,HuangSCPMA}. Along this line, there has been accumulated interest in studying the holographic $s$-wave \cite{Arean,SonnerWithers,KuangLiuWang,Zeng2013,Amado2014}, $p$-wave \cite{Zeng2011,RogatkoWysokinski,LvPLB2020,LaiHPJPWaveSolition,Nie-Zeng} and $s+p$-wave \cite{AriasLandea} superfluid models.

However, all the studies mentioned above concerning the holographic superconductor/superfluid models are mainly based on the ground state since it is the first state to condense. As a matter of fact, in condensed matter physics, the physical system is not necessarily in equilibrium, but may remain the excited metastable states which manifest themselves in the hysteresis, superheating and supercooling phenomena, in the paramagnetic Meissner effect, in the jumps of magnetization, and in other peculiarities of the mesoscopic samples behavior, observed experimentally \cite{Zharkov}. Thus, considering the importance of the excited states for superconducting materials in condensed matter systems \cite{Peeters2000,Vodolazov2002,RMP2004,RMP2011}, Wang \emph{et al.} constructed the numerical solutions of holographic s-wave superconductors with the excited states and noted that the excited state has a lower critical temperature than the corresponding ground state \cite{WangYQ2020}. Extending the investigation to the backreaction in the Einstein gravity \cite{WangLLZEPJC2021} and 4D Einstein-Gauss-Bonnet gravity \cite{PanJie}, the authors observed that there exist the additional peaks in the imaginary and real parts of the conductivity, and the number of peaks is equal to the number of nodes of the $n$-th excited state. Analyzing the non-equilibrium dynamical transition process between excited states of holographic s-wave superconductors, Li \emph{et al.} pointed out the dynamical formation of the excited states as the intermediate states during the relaxation from the normal state to the ground state \cite{LiWWZJHEP}. In order to back up numerical results, we proposed the improved variational method for the Sturm-Liouville eigenvalue problem to investigate the excited states of the holographic superconductors, which shows that this generalized Sturm-Liouville method is very powerful to study the properties of the phase transition with the excited states \cite{QiaoEHS,OuYangliangSCPMA}. Considering the increasing interest in study of the excited states by holography \cite{XiangZWCTP,BGKuangPLB,ZhangZPJNPB,NguyenJHEP,XiangZW,WangNPB2023}, in this work we will extend the investigation to the holographic superfluid with the excited states, which has not been constructed as far as we know. For simplicity, we focus on the probe approximation where the backreaction of matter fields on the spacetime metric is neglected. We will find that the combination of the excited state and the superfluid velocity provides richer physics in the scalar condensates and conductivity in the holographic superfluid model.

This work is organized as follows. In Sec. II we will construct the holographic s-wave superfluid model in the AdS black hole background. In Sec. III we will investigate the condensates of the scalar field and the phase transition in the holographic superfluid both for the ground state and excited states. In Sec. IV we will calculate the electrical conductivity of the holographic superfluid with the excited states. We will conclude in the last section with our main results.

\section{Description of the holographic superfluid model}

We will work with the action containing a Maxwell field and a charged complex
scalar field coupled via a generalized Lagrangian
\begin{eqnarray}\label{System}
S=\int
d^{d+1}x\sqrt{-g}\left\{\frac{1}{2\kappa^{2}}\left[R+\frac{d(d-1)}{L^{2}}\right]
-\frac{1}{4}F_{\mu\nu}F^{\mu\nu}-|\nabla\psi-iqA\psi|^{2}
-m^2|\psi|^2\right\},
\end{eqnarray}
with the $(d+1)$-dimensional gravitational constant $\kappa^{2}=8\pi G_{d+1}$ and AdS radius $L$. Here $F_{\mu\nu}=\nabla_{\mu}A_{\nu}-\nabla_{\nu}A_{\mu}$ is the strength tensor of the gauge field $A_{\mu}$, and $\psi$ is a scalar field with charge $q$ and mass $m$. Since we focus on the probe approximation, we will obtain the $(d+1)$-dimensional planar Schwarzschild-AdS black hole in the form
\begin{eqnarray}\label{bhmetric}
ds^2&=&-f(r)dt^{2}+\frac{dr^2}{f(r)}+r^{2}dx_{i}dx^{i},
\end{eqnarray}
where the metric coefficient $f(r)=r^{2}(1-r_{+}^{d}/r^{d})/L^{2}$ with the black hole horizon $r_{+}$. The corresponding Hawking temperature is $T=dr_{+}/(4\pi L^{2})$, which can be interpreted as the temperature of CFT.

In order to construct the holographic superfluid model in the AdS black hole background, we will take the ansatz of the matter fields as
\begin{eqnarray}
\psi=\psi(r),~~A_{\mu}=(A_{t}(r), 0, A_{x}(r), ...),
\end{eqnarray}
where $\psi$, $A_{t}$ and $A_{x}$ are real functions of $r$ only. Thus, we have the equations of motion
\begin{eqnarray}\label{equtions}
\psi^{\prime\prime}+\bigg(\frac{d-1}{r}+\frac{f^{\prime}}{f}\bigg)\psi^{\prime}+\bigg(\frac{q^{2}A_{t}^{2}}{f^{2}}-\frac{q^{2}A_{x}^{2}}{r^{2}f}-\frac{m^{2}}{f}\bigg)\psi&=&0,\notag \\
A^{\prime\prime}_{t}+\frac{d-1}{r}A^{\prime}_{t}-\frac{2q^{2}\psi^{2}}{f}A_{t}&=&0,\notag \\
A^{\prime\prime}_{x}+\bigg(\frac{d-3}{r}+\frac{f^{\prime}}{f}\bigg)A^{\prime}_{x}-\frac{2q^{2}\psi^{2}}{f}A_{x}&=&0,  \label{bhpsi}
\end{eqnarray}
where the prime denotes differentiation in $r$.

With the appropriate boundary conditions, we will solve the equations of motion (\ref{equtions}) numerically by using the shooting method \cite{HartnollPRL}. At the horizon $r=r_{+}$, the fields $\psi$ and $A_{x}$ are regular but $A_{t}$ satisfies $A_{t}(r_{+})=0$. Near the AdS boundary $(r\rightarrow \infty)$, the asymptotic behaviors of the solutions are
\begin{eqnarray}
\psi=\frac{\psi_{-}}{r^{\triangle_{-}}}+\frac{\psi_{+}}{r^{\triangle_{+}}}, ~~~~A_{t}=\mu-\frac{\rho}{r^{d-2}},~~~~A_{x}=S_{x}-\frac{J_{x}}{r^{d-2}},
\end{eqnarray}
with the conformal dimension of the scalar operator $\triangle_{\pm}=(d\pm\sqrt{d^{2}+4m^{2}L^{2}})/2$ dual to the bulk scalar field. Here, $\mu$ and $S_{x}$ are interpreted as the chemical potential and superfluid velocity, while $\rho$ and $J_{x}$ are identified as the charge density and current in the dual field theory, respectively. From the gauge/gravity duality, provided $\triangle_{-}$ is larger than the unitarity bound, both $\psi_{-}$ and $\psi_{+}$ are normalizable and can be used to define the dual scalar operator, $\psi_{-}=\langle \mathcal{O}_{-}\rangle$, $\psi_{+}=\langle \mathcal{O}_{+}\rangle$ \cite{HartnollPRL,HartnollJHEP12}, respectively. We will impose boundary condition $\psi_{-}=0$ since the other choice will not qualitatively modify our results. Thus, we set $\mathcal{O}=\mathcal{O}_{+}$ and $\triangle=\triangle_{+}$ for clarity. In numerics, we do integration from the horizon out to the infinity and choose two independent parameters of the fields $\psi$, $A_{x}$ and $A_{t}$ at the horizon as shooting parameters to match the source free condition, which leads to a divergence in the functions. So we iterate the procedure, fine-tuning our initial conditions at the horizon such that the regular solutions are found. We discretize the coupled differential equations (\ref{equtions}) by using a finite difference method, and adjust the mesh spacing until we achieve stability for our code. In this work, the relative error for the numerical solutions is estimated to be below $10^{-4}$.

It should be noted that, in order to study the thermodynamical stability of the superconducting as well as normal phases, we will compute the grand potential $\Omega=-T\mathcal{S}_{os}$ of the bound state. According to the action (\ref{System}), in the probe limit we can express the Euclidean on-shell action $\mathcal{S}_{os}$ as
\begin{eqnarray}
\mathcal{S}_{os}&=&\int d^{d}x\left(\frac{1}{2}r^{d-1}A_{t}A_{t}^{\prime}-\frac{1}{2}r^{d-3}fA_{x}A_{x}^{\prime}-r^{d-1}f\psi\psi^{\prime}\right)|_{r=\infty}+\int d^{d+1}x\left(q^{2}r^{d-3}A_{x}^{2}-\frac{q^{2}r^{d-1}A_{t}^{2}}{f}\right)\psi^{2}\notag \\
&=&\frac{V_{d-1}}{T}\left[\frac{(d-2)}{2}\mu\rho-\frac{(d-2)}{2}S_{x}J_{x}+\int q^{2}r^{d-3}\psi^{2}\left(A_{x}^{2}-\frac{r^{2}A_{t}^{2}}{f}\right)dr\right],
\end{eqnarray}
with the integration $\int d^{d}x=V_{d-1}/T$. Therefore, the grand potential in the superfluid phase is given by
\begin{eqnarray}
\frac{\Omega_{S}}{V_{d-1}}=-\frac{T\mathcal{S}_{os}}{V_{d-1}}=-\frac{(d-2)}{2}\mu\rho+\frac{(d-2)}{2}S_{x}J_{x}-\int_{r_{+}}^{\infty} q^{2}r^{d-3}\psi^{2}\left(A_{x}^{2}-\frac{r^{2}A_{t}^{2}}{f}\right)dr.
\end{eqnarray}
For the normal phase, i.e., $\psi=0$, we get the grand potential $\Omega_{N}/V_{d-1}=-(d-2)\mu\rho/2$ in this case.

\section{Condensates of the scalar field}

In Ref. \cite{AreanJHEP}, the authors constructed the holographic superfluid in AdS$_{d+1}$ with $d=3$ and $4$ in the probe limit for various masses of the scalar field and obtained the rich phase structure of the system in the ground state. Now we
generalize the investigation on the phase transition in the holographic superfluid with the excited states which, as far as we know, has not been explored yet.

It is interesting to note that, from the equations of motion (\ref{equtions}), we have the useful scaling symmetries and the transformation of relevant quantities
\begin{eqnarray}
&&r\rightarrow\lambda r\,,\hspace{0.5cm}(t, x^{i})\rightarrow\frac{1}{\lambda}(t, x^{i})\,,\hspace{0.5cm}(q, \psi)\rightarrow
(q, \psi)\,,\hspace{0.5cm}(A_{t},A_{x})\rightarrow\lambda(A_{t},A_{x})\,,\nonumber \\
&&\psi_{\pm}\rightarrow\lambda^{\triangle_{\pm}}\psi_{\pm}\,,\hspace{0.5cm}
(T,\mu,S_x)\rightarrow\lambda(T,\mu,S_x)\,,\hspace{0.5cm}
(\rho,J_x)\rightarrow\lambda^{d-1}(\rho,J_x)\,,
\label{PWSLsymmetry}
\end{eqnarray}
with the positive number $\lambda$. Thus, we will take advantage of the scaling symmetries to set $q=1$ and $r_{+}=1$ when performing numerical calculations. For convenience, we will take the coordinate transformation $r\rightarrow z=r_{+}/r$. Just as in \cite{AreanJHEP}, we focus on the cases of $d=3$ and $4$ for concreteness.

\subsubsection{The case of $d=3$}

\begin{figure}[ht]
\includegraphics[scale=0.62]{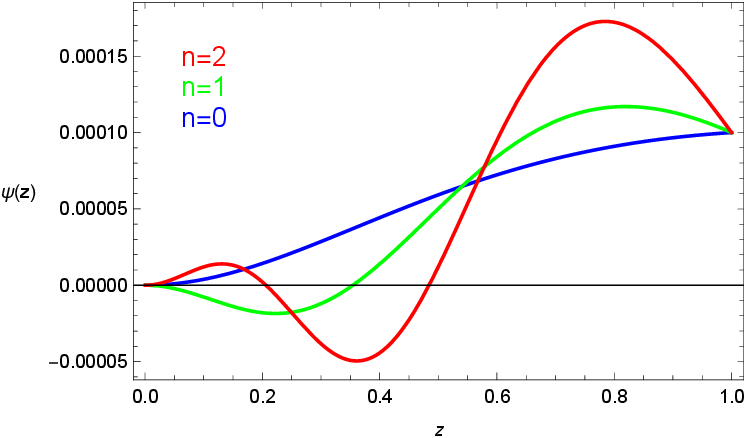}\hspace{0.4cm}%
\includegraphics[scale=0.62]{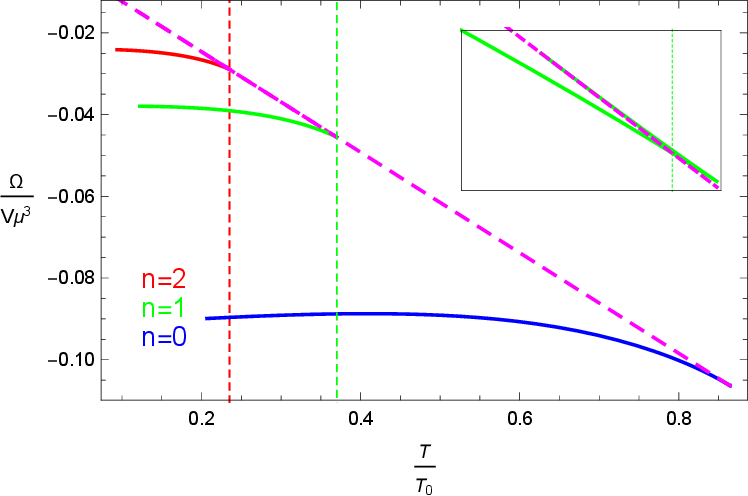}\hspace{0.4cm}%
\caption{\label{Psiz} (Color online) The scalar field $\psi(z)$ as a function of the radial coordinate $z$ outside the horizon (left) and the grand potential $\Omega$ as a function of the temperature $T$ (right) for the fixed mass of the scalar field $m^{2}L^{2}=-2$ and superfluid velocity $S_{x}/\mu=0.24$ in the case of $d=3$. In each panel, the blue, green and red solid lines denote the ground $(n=0)$, first $(n=1)$ and second $(n=2)$ states, respectively. For the right panel, the magenta dotted line corresponds to the normal phase and the vertical line represents the critical temperature (critical chemical potential) of the first-order phase transition. }
\end{figure}

It is well known that the excited states correspond to the bulk solutions for which the scalar field changes sign along the radial direction. In the left panel of Fig. \ref{Psiz}, we present the distribution of the scalar field $\psi(z)$ as a function of $z$ for the fixed scalar mass $m^{2}L^{2}=-2$ and superfluid velocity $S_{x}/\mu=0.24$ by setting the initial condition $\psi(1)=0.0001$, which shows that the excited states are characterized by the number of nodes of the scalar field in the radial direction and the ground state refers to the scalar field without nodes. In Ref. \cite{WangYQ2020}, Wang \emph{et al.} revealed some universal properties in the holographic superconductors with the excited states, i.e., the excited state always has a lower critical temperature (larger critical chemical potential) than the corresponding ground state, and there exist additional poles in Im[$\sigma(\omega)$] and delta functions in Re[$\sigma(\omega)$] arising at low temperature. In this work, we will observe that these properties to be the same even in the holographic superfluid models with the excited states. On the other hand, in the right panel of Fig. \ref{Psiz}, we give the grand potential as a function of the temperature for the fixed scalar mass $m^{2}L^{2}=-2$ and superfluid velocity $S_{x}/\mu=0.24$, which shows that the excited state always has a larger grand potential than the corresponding ground state and thus metastable. This can be used to back up the findings obtained in Ref. \cite{WangYQ2020} that the excited states of the holographic superconductors could be related to the metastable states of the mesoscopic superconductors.

\begin{figure}[ht]
\includegraphics[scale=0.41]{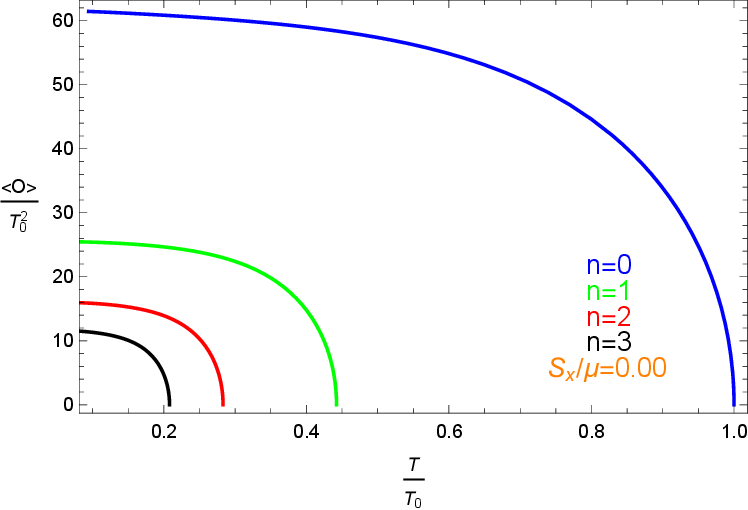}\hspace{0.4cm}%
\includegraphics[scale=0.41]{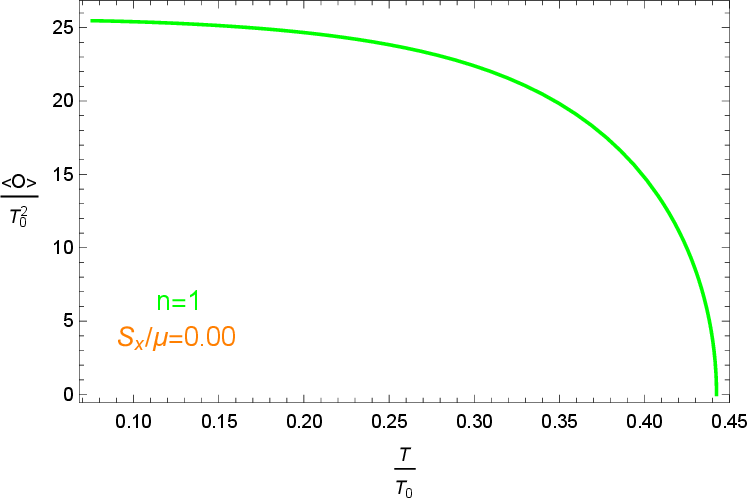}\hspace{0.4cm}%
\includegraphics[scale=0.42]{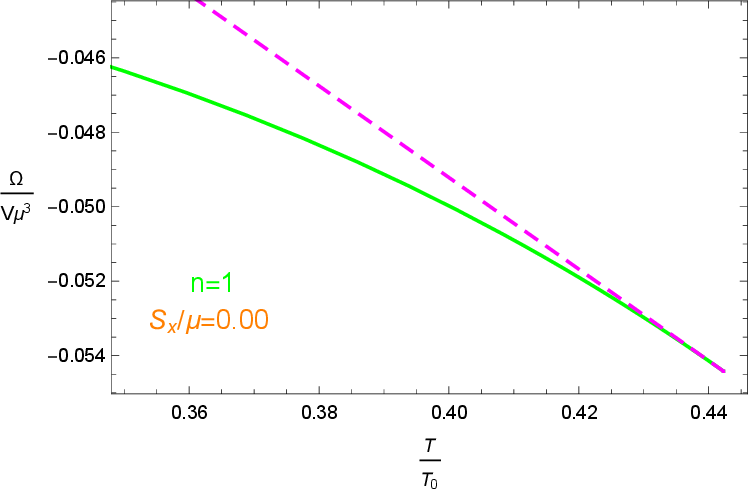}\hspace{0.4cm}%
\includegraphics[scale=0.41]{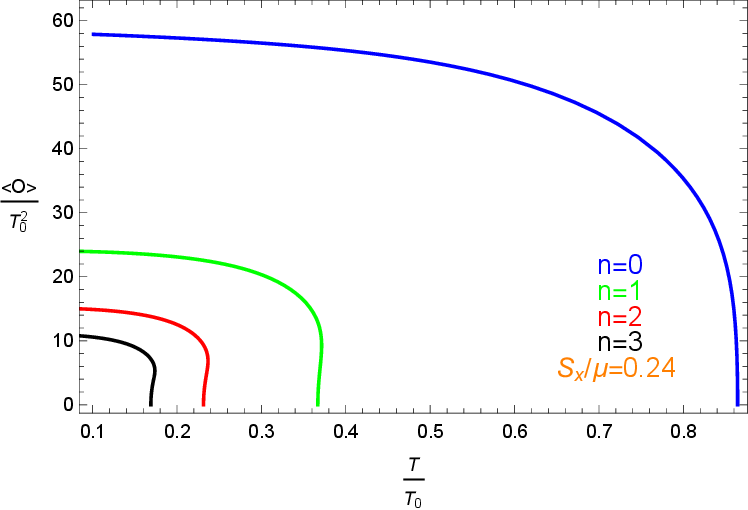}\hspace{0.4cm}%
\includegraphics[scale=0.41]{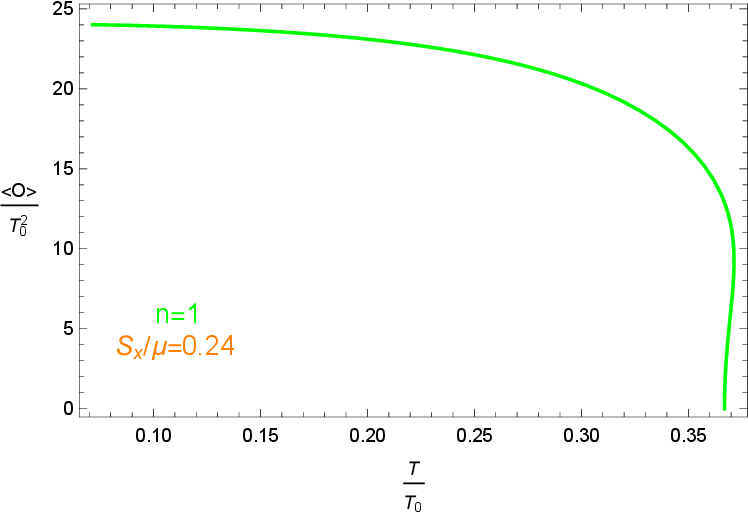}\hspace{0.4cm}%
\includegraphics[scale=0.42]{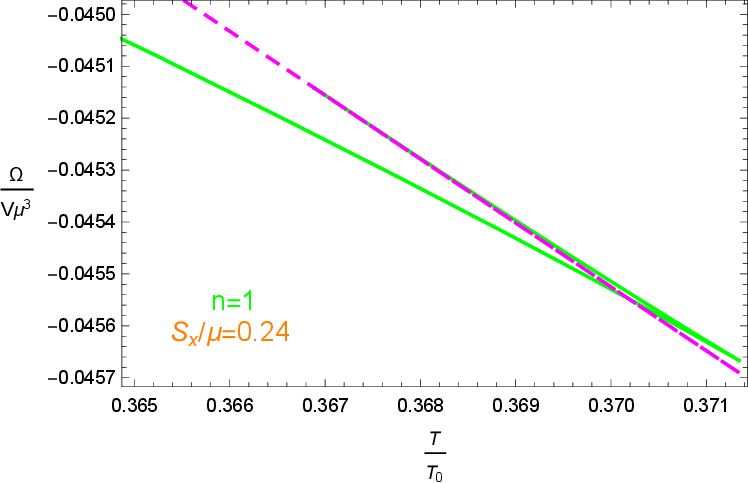}\hspace{0.4cm}%
\includegraphics[scale=0.41]{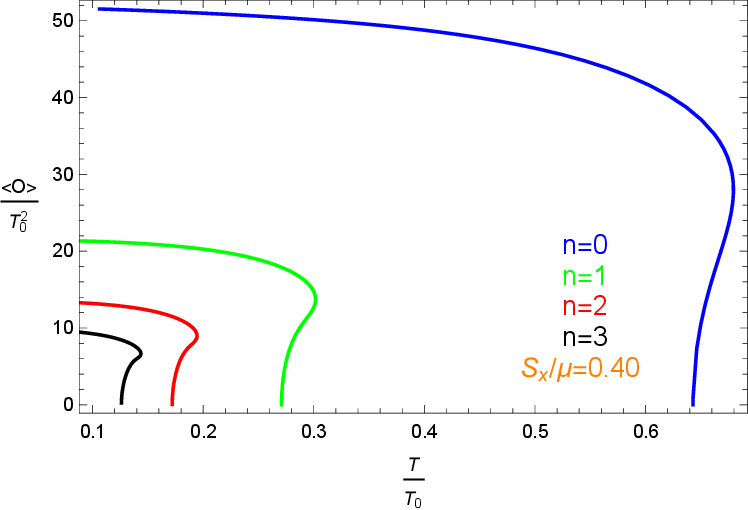}\hspace{0.4cm}%
\includegraphics[scale=0.41]{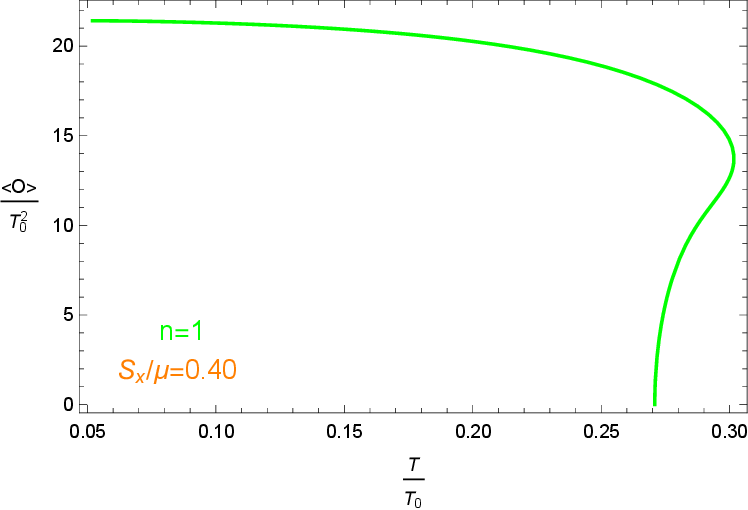}\hspace{0.4cm}%
\includegraphics[scale=0.42]{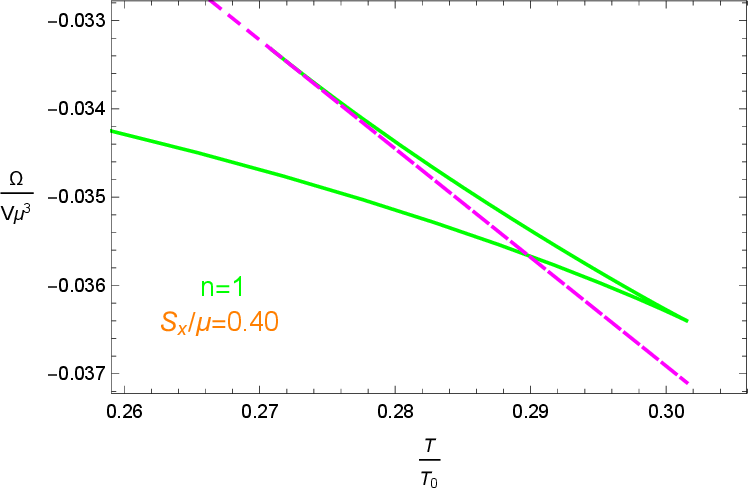}\hspace{0.4cm}%
\caption{ (Color online) The condensate and grand potential as a function of the temperature with the fixed mass of the scalar field $m^{2}L^{2}=-2$ for different values of $S_{x}/\mu$ from the ground state to the third excited state in the case of $d=3$. For the left three panels, the four lines in each panel from top to bottom correspond to the ground ($n=0$), first ($n=1$), second ($n=2$) and third ($n=3$) states, respectively. For the middle three panels, the line in each panel corresponds to the first state $n=1$. For the right three panels, the two lines in each panel correspond to the first state $n=1$ (green solid) and the normal phase (magenta dotted) respectively. }\label{4dbhCondMLF2}
\end{figure}

In Fig. \ref{4dbhCondMLF2}, we plot the condensate and the corresponding grand potential as a function of the temperature $T$ with $m^{2}L^{2}=-2$ from the ground state to the third excited state in the case of $d=3$ for different values of $S_{x}/\mu$, i.e., $S_{x}/\mu=0.00$, $0.24$ and $0.40$, where the critical temperature $T_{0}$ corresponds to the case of $n=0$ and $S_{x}/\mu=0.00$. Obviously, the critical temperature $T_{c}$ decreases as the number of nodes $n$ or the superfluid velocity $S_{x}/\mu$ increases, which indicates that the higher excited state or larger superfluid velocity will make the scalar hair more difficult to be developed. In the case of vanishing or small superfluid velocity, such as $S_{x}/\mu=0.00$ and $S_{x}/\mu=0.24$ with $n=0$, the holographic superfluid phase transition is always the second-order one with the mean field critical exponent $1/2$, i.e., $\langle {\cal O}\rangle\sim (1-T/T_{c})^{1/2}$, which agrees well with the results of holographic superconductors both for the ground state and excited states in Ref. \cite{WangYQ2020}. However, the superfluid phase transition will change from the second order to the first order when the superfluid velocity increases, for example $S_{x}/\mu=0.24$ with $n=1$, $2$, $3$ and $S_{x}/\mu=0.40$ both for the ground state and excited states, which can be derived from the grand potential characterized by the typical swallowtails in the bottom two panels of the rightmost column for the case of $n=1$. Obviously, there exists a turning point $S_{x}/\mu$ where the transition switches from the second order to the first order, which depends on the number of nodes $n$. As a matter of fact, this turning point $S_{x}/\mu$ also depends on the scalar mass $m^{2}L^{2}$.

\begin{figure}[ht]
\includegraphics[scale=0.68]{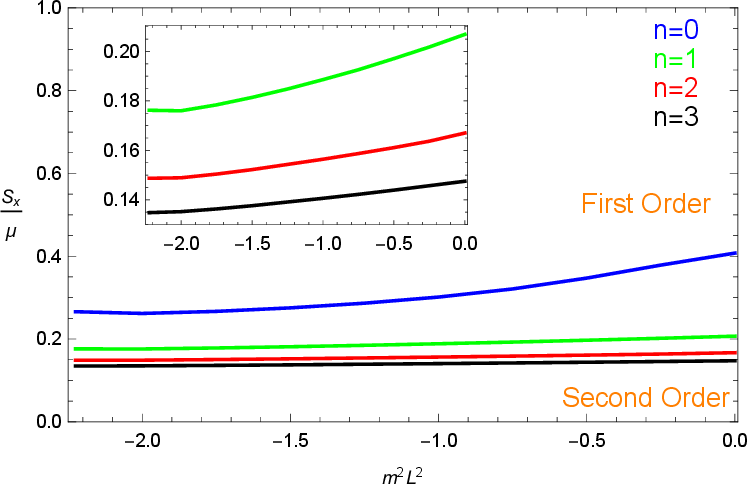}\hspace{0.4cm}%
\caption{ (Color online) The translating superfluid velocity $S_{x}/\mu$ from the second to the first order as a function of the the scalar mass $m^{2}L^{2}$ for the ground ($n=0$, blue), first ($n=1$, green), second ($n=2$, red) and third ($n=3$, black) states, respectively. }\label{4dbhSVML}
\end{figure}

In order to further investigate the effect of the number of nodes and the scalar mass on the turning point from the second to the first order, we plot the translating superfluid velocity $S_{x}/\mu$ as a function of mass of the scalar field $m^{2}L^{2}$ from the ground state to the third excited state in Fig. \ref{4dbhSVML}. It is shown that, both for the ground state and excited states, the translating superfluid velocity $S_{x}/\mu$ increases monotonously with the increasing scalar mass $m^{2}L^{2}$. On the other hand, for the fixed mass of the scalar field $m^{2}L^{2}$, we observe that the translating superfluid velocity $S_{x}/\mu$ decreases as the number of nodes $n$ increases. Thus, the higher excited state or smaller mass of the scalar field  makes it easier for the emergence of translating point from the second-order transition to the first-order one.

\begin{table}[ht]
\begin{center}
\caption{The critical chemical potential $\mu_{c}$ with the fixed mass of the scalar field $m^{2}L^{2}=-2$ for different values of $S_{x}/\mu$ from the ground state to the sixth excited state in the case of $d=3$. }
\begin{tabular}{|c|c|c|c|c|c|c|c|}
  \hline
  $n$ & 0 & 1 & 2 & 3 & 4 & 5& 6\\
  \hline
~~$S_{x}/\mu=0.00$~~&~~$4.064$~~&~~$9.188$~~&~~$14.357$~~&~~$19.539$~~&~~$24.726$~~&~~$29.916$~~ &~~$35.106$~~ \\
  \hline
~~$S_{x}/\mu=0.24$~~&~~$4.702$~~&~~$10.977$~~&~~$17.277$~~&~~$23.586$~~&~~$29.900$~~&~~$36.216$~~ &~~$42.533$~~  \\
  \hline
~~$S_{x}/\mu=0.40$~~&~~$6.074$~~&~~$14.021$~~&~~$21.999$~~&~~$29.987$~~&~~$37.978$~~ &~~$45.972$~~ &~~$53.968$~~ \\
 \hline
\end{tabular}
\end{center}
\end{table}

As a matter of fact, we also want to know the influence of the superfluid velocity on the critical chemical potential $\mu_{c}$. For the first-order phase transition, just as in the right panel of Fig. 1, we calculate the grand potential $\Omega$ for the superconducting phase and the normal phase, and mark the location of the critical chemical potential with a vertical dotted line in the same color as the grand-potential curve. Thus, in Table 1 we give the critical chemical potential $\mu_{c}$ with the fixed mass of the scalar field $m^{2}L^{2}=-2$ for different values of $S_{x}/\mu$ from the ground state to the sixth excited state in the case of $d=3$, which shows that, regardless of $S_{x}/\mu$, the excited state always has a larger critical chemical potential than the corresponding ground state. Fitting the relation between $\mu_{c}$ and $n$ by using the numerical results, we obtain
\begin{eqnarray}\label{SWaveMuc}
\mu_{c}\approx
\left\{
\begin{array}{rl}
5.177n+4.026, &   \ S_{x}/\mu=0.00,\\ \\
6.307n+4.678, &   \ S_{x}/\mu=0.24,\\ \\
7.984n+6.047, &   \ S_{x}/\mu=0.40,
\end{array}\right.
\end{eqnarray}
which shows that, although the underlying mechanism is still unclear, the critical chemical potential $\mu_{c}$ becomes evenly spaced for the number of nodes $n$, and the difference of $\mu_{c}$ between the consecutive states increases as the superfluid velocity $S_{x}/\mu$ increases. It should be noted that this conclusion still holds in higher dimensions.

\subsubsection{The case of $d=4$}

Now we move to the case of $d=4$. For the small mass scale and large mass scale, for example $m^{2}L^{2}=-15/4$ and $m^{2}L^{2}=0$ just as in \cite{AreanJHEP}, our findings are similar to those obtained in the case of $d=3$. Concretely, the critical temperature $T_{c}$ decreases with the increase of $n$ or $S_{x}/\mu$, which implies that the higher excited state or larger superfluid velocity makes the condensate harder. And the number of nodes $n$ can change the order of the phase transition in the holographic superfluid system and the higher excited state results in smaller superfluid velocity for the turning point, where the second-order phase transition changes into the first order. However, the intermediate mass scale will exhibit a very interesting and different feature. So we concentrate on this case.

\begin{figure}[ht]
\includegraphics[scale=0.41]{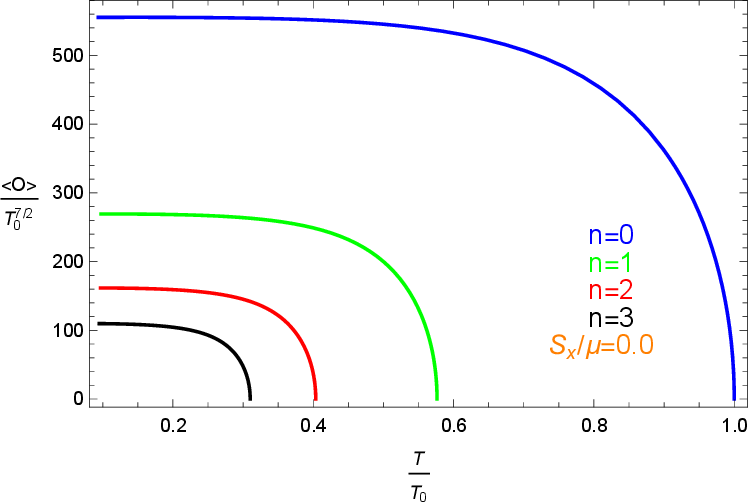}\hspace{0.4cm}%
\includegraphics[scale=0.41]{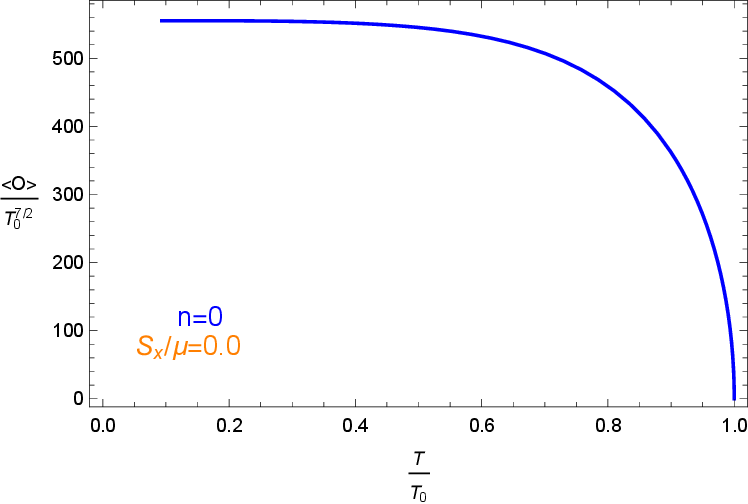}\hspace{0.4cm}%
\includegraphics[scale=0.42]{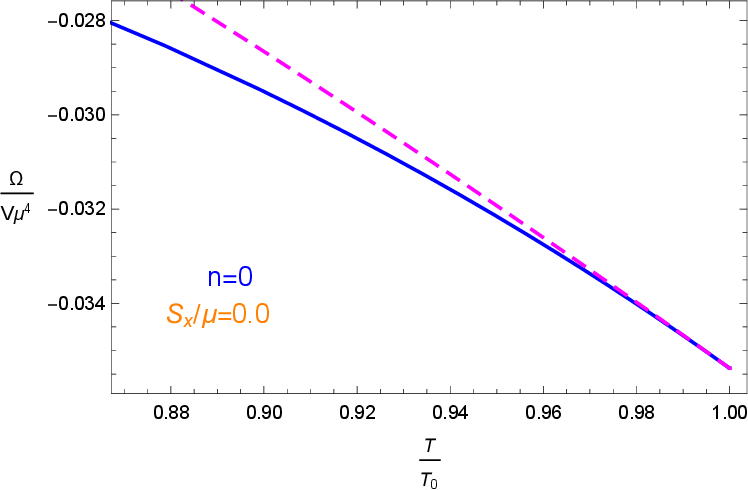}\hspace{0.4cm}%
\includegraphics[scale=0.41]{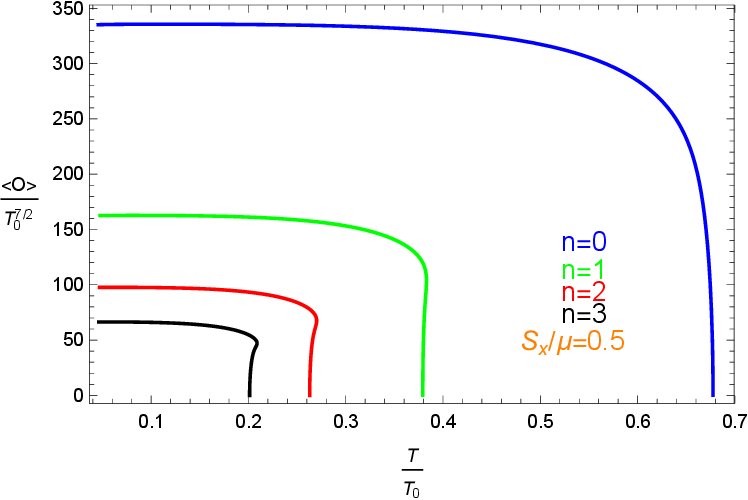}\hspace{0.4cm}%
\includegraphics[scale=0.41]{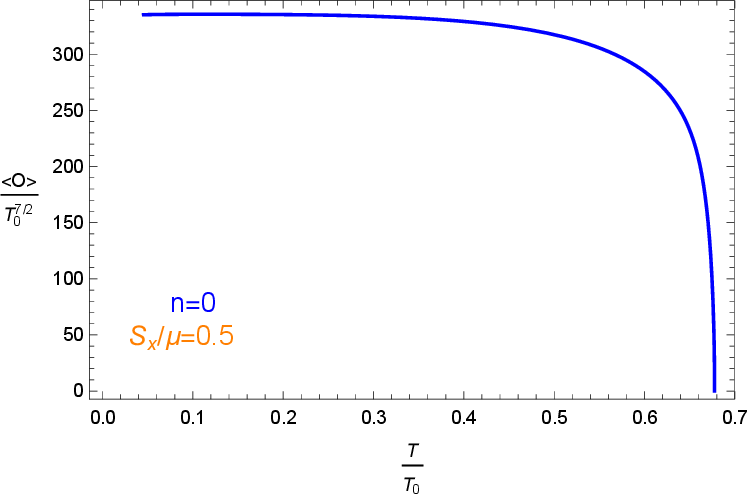}\hspace{0.4cm}%
\includegraphics[scale=0.42]{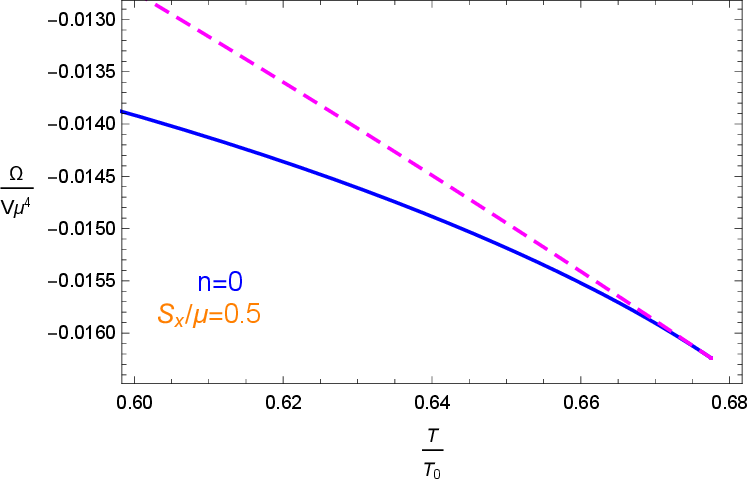}\hspace{0.4cm}%
\includegraphics[scale=0.41]{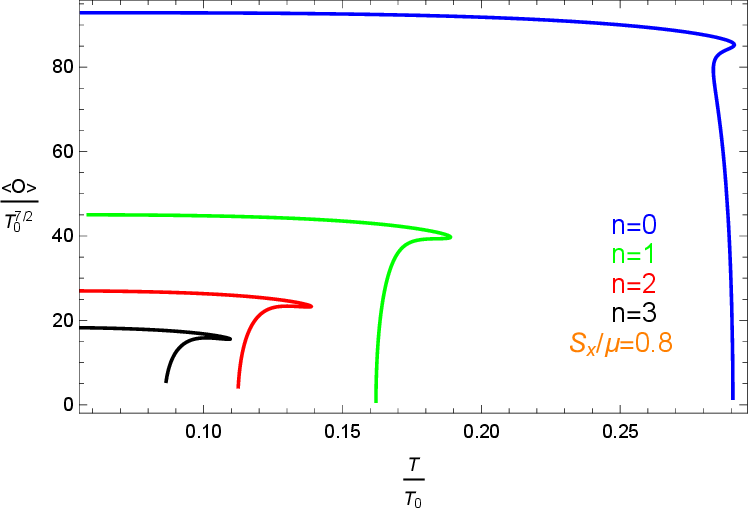}\hspace{0.4cm}%
\includegraphics[scale=0.41]{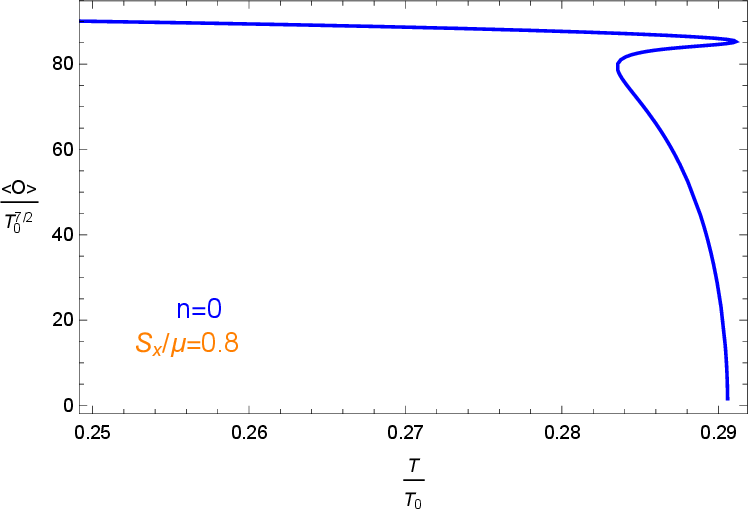}\hspace{0.4cm}%
\includegraphics[scale=0.42]{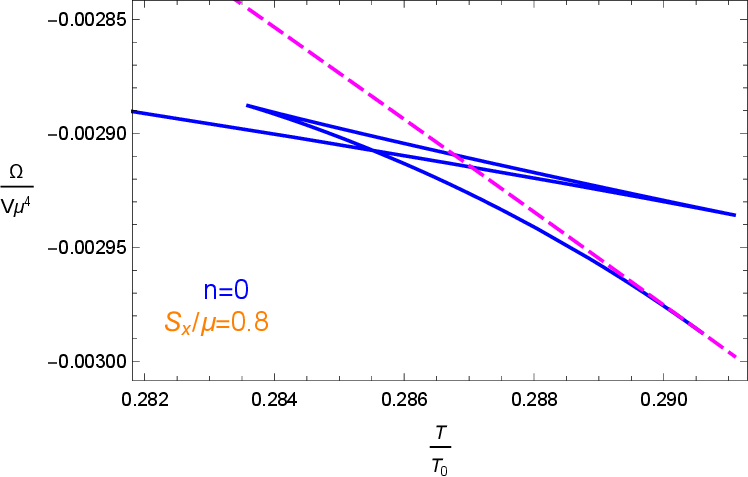}\hspace{0.4cm}%
\caption{ (Color online) The condensate and grand potential as a function of the temperature with the fixed mass of the scalar field $m^{2}L^{2}=-7/4$ for different values of $S_{x}/\mu$ from the ground state to the third excited state in the case of $d=4$. For the left three panels, the four lines in each panel from top to bottom correspond to the ground ($n=0$), first ($n=1$), second ($n=2$) and third ($n=3$) states, respectively. For the middle three panels, the line in each panel corresponds to the ground state $n=0$. For the right three panels, the two lines in each panel correspond to the ground state $n=0$ (blue solid) and the normal phase (magenta dotted) respectively. }\label{bhCondML175}
\end{figure}

Considering the intermediate mass scale, for example $m^{2}L^{2}=-7/4$, we plot the condensate and the corresponding grand potential as a function of the temperature $T$ from the ground state to the third excited state for $S_{x}/\mu=0.0$, $0.5$ and $0.8$ in Fig. \ref{bhCondML175}. When the superfluid velocity is small, for example $S_{x}/\mu=0.0$ with $n=0$, $1$, $2$, $3$ and $S_{x}/\mu=0.5$ with $n=0$, the holographic superfluid phase transition is found to be second order and the condensate satisfies $\langle {\cal O}\rangle\sim (1-T/T_{c})^{1/2}$ near the critial temperature. When the superfluid velocity is large enough, such as $S_{x}/\mu=0.8$, we observe that the scalar operator ${\cal O}$ becomes multivalued for given temperatures and the Cave of Winds appears, which is associated with the emergence of a second superfluid phase and determined via the corresponding grand potential shown in the bottom panel of the rightmost column in Fig. \ref{bhCondML175}. However, this phenomenon only occurs in the ground state, i.e., $n=0$. If the number of nodes $n$ increases, the Cave of Winds disappears but the first-order phase transition occurs, for example $S_{x}/\mu=0.8$ with $n=1$, $2$ and $3$ in this figure. As a matter of fact, the other choices of $S_{x}/\mu$ will not change our results qualitatively. Thus, it is interesting to note that the excited state will hinder the appearance of the Cave of Winds.

\section{Conductivity}

Now we want to calculate the conductivity of the holographic superfluid with the excited states by considering the perturbed Maxwell field $\delta A_{y}=e^{-i\omega t}A_{y}(r)dy$. So the equation of motion takes the form
\begin{eqnarray}\label{CDequation}
A^{\prime\prime}_{y}+\left(\frac{d-3}{r}+\frac{f^{\prime}}{f}\right)A^{\prime}_{y}+\bigg(\frac{\omega^{2}}{f^{2}}-\frac{2q^{2}\psi^{2}}{f}\bigg)A_{y}=0,
\end{eqnarray}
where the ingoing wave boundary conditions at the horizon is $A_{y}\varpropto f^{-i\omega/(dr_{+})}$. For concreteness, we will restrict our study to $d=3$ and $m^{2}L^{2}=-2$ since the other choices will not change our results qualitatively. Thus, for $d=3$ dimension, the behavior in the asymptotic region ($r\rightarrow\infty$) reads
\begin{eqnarray}
A_{y}=A_{y}^{(0)}+\frac{A_{y}^{(1)}}{r}+\cdots.
\end{eqnarray}
According to the gauge/gravity duality, we can express the conductivity of the dual superconductor as
\begin{eqnarray}
\sigma=-\frac{iA_{y}^{(1)}}{\omega A_{y}^{(0)}}.
\end{eqnarray}
For different values of the number of nodes $n$ and the superfluid velocity $S_{x}/\mu$, we focus on the scalar operator $\mathcal{O}=\mathcal{O}_{+}$ and give the conductivity by solving the Maxwell equation (\ref{CDequation}) numerically.

\label{bhAx}
\begin{figure}[ht]
\includegraphics[scale=0.41]{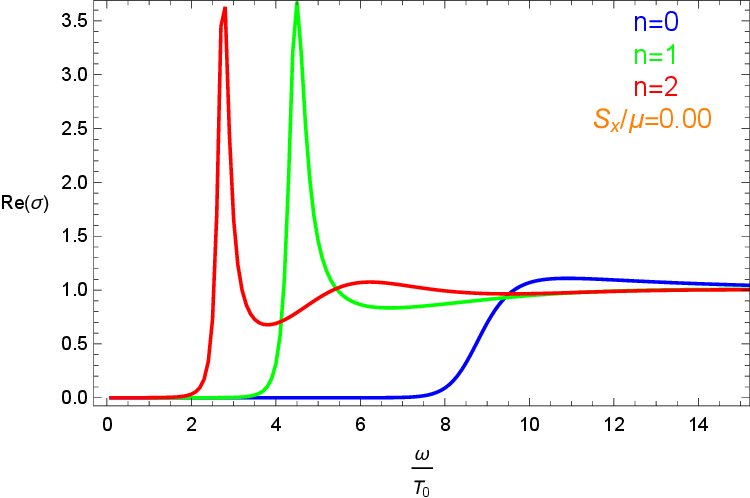}\hspace{0.4cm}%
\includegraphics[scale=0.41]{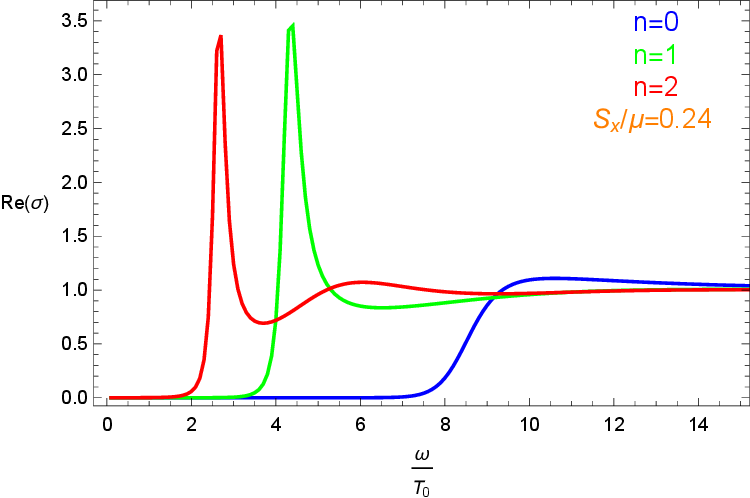}\hspace{0.4cm}%
\includegraphics[scale=0.41]{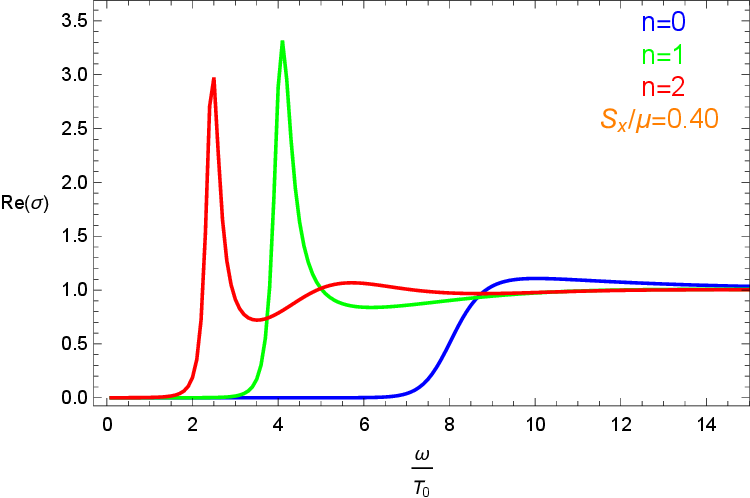}\hspace{0.4cm}%
\includegraphics[scale=0.41]{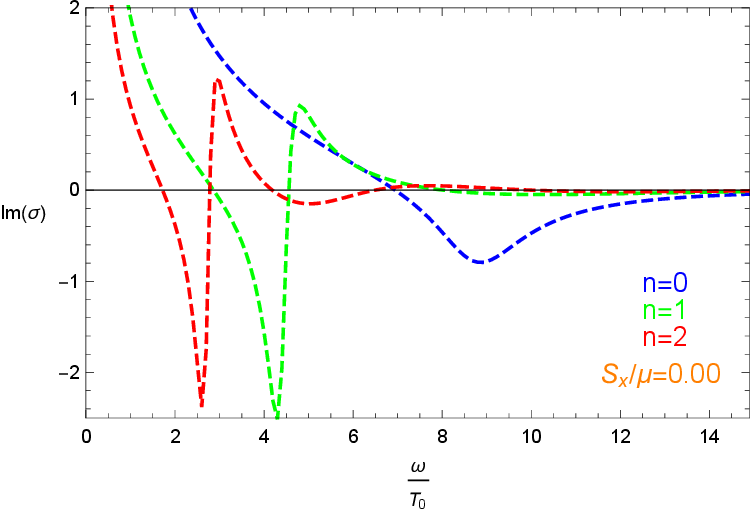}\hspace{0.4cm}%
\includegraphics[scale=0.41]{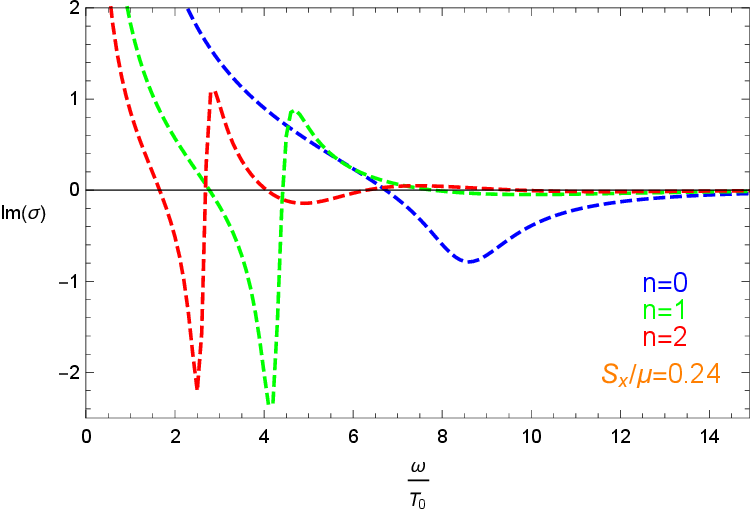}\hspace{0.4cm}%
\includegraphics[scale=0.41]{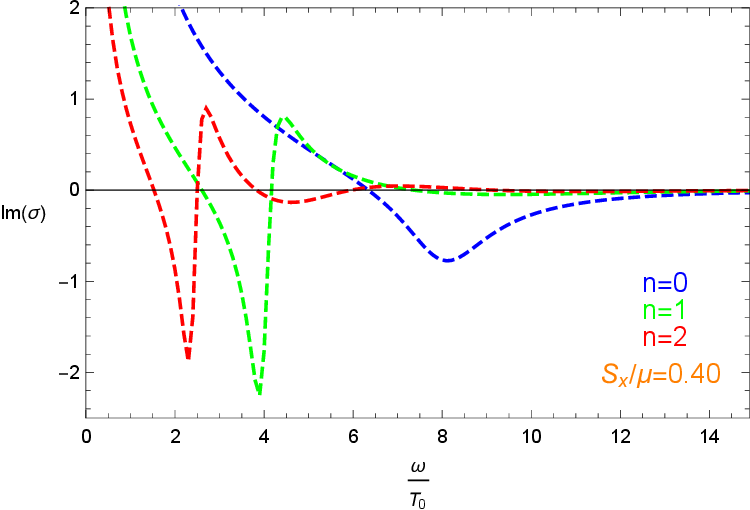}\hspace{0.4cm}%
\caption{ (Color online) The conductivity in the holographic superfluid with the fixed mass of the scalar field $m^{2}L^{2}=-2$ for different values of $S_{x}/\mu$ from the ground state to the second excited state where the solid line and dashed line represent the real part Re$(\sigma)$ and imaginary part Im$(\sigma)$ of the conductivity. In each panel, the blue, green and red lines denote the ground ($n=0$), first ($n=1$) and second ($n=2$) states, respectively. }\label{ConductivityZF2}
\end{figure}

In Fig. \ref{ConductivityZF2}, we plot the conductivity $\sigma(\omega)$ in the holographic superfluid with the fixed mass of the scalar field $m^{2}L^{2}=-2$ for different values of $S_{x}/\mu$, i.e., $S_{x}/\mu=0.00$, $0.24$ and $0.40$, from the ground state to the second excited state at temperatures $T/T_{0}\simeq0.15$ with the critical temperature $T_{0}$ in the case of $n=0$ and $S_{x}/\mu=0.00$, just as in Ref. \cite{Arean}. We can see clearly that, for the excited states, there exist additional poles in Im[$\sigma(\omega)$] and delta functions in Re[$\sigma(\omega)$] arising at low temperature, which is similar to the excited states of holographic superconductors \cite{WangYQ2020}. However, both for the ground state and excited states, there exists a gap in the conductivity and we can define the gap frequency $\omega_{g}$ as the frequency minimizing Im[$\sigma(\omega)$] \cite{Horowitz2008}. Obviously, regardless of either the ground state or excited state, it is shown that increasing superfluid velocity $S_{x}/\mu$ decreases $\omega_{g}/T_{0}$, which can be used to back up the numerical finding as shown in figure 7 of Ref. \cite{Arean} that $\omega_{g}/T_{c}$ decreases as the current increases for the ground state. On the other hand, for a fixed superfluid velocity $S_{x}/\mu$, as we increase the number of nodes $n$, the gap frequency $\omega_{g}/T_{0}$ decreases, which tells us that the higher excited state results in the larger deviation from the expected relation in the gap frequency $\omega_{g}/T_{c}\approx8$ obtained in \cite{Horowitz2008}.

\section{Conclusion}

We have presented a novel family of solutions of the holographic superfluid model with the excited states in order to understand the excited state in superconducting materials in condensed matter systems by holography. In the probe limit, we found that the excited state always has a larger grand potential than the corresponding ground state and thus metastable, and the critical temperature decreases as the number of nodes $n$ or the superfluid velocity $S_{x}/\mu$ increases, which implies that the higher excited state or larger superfluid velocity will make it harder for the scalar hair to form. For our holographic superfluid model, we observed that there exists a turning point $S_{x}/\mu$ where the transition switches from the second order to the first order, which shows that the translating superfluid velocity $S_{x}/\mu$ decreases as the number of nodes $n$ increases or the scalar mass $m^{2}L^{2}$ decreases. This means that the higher excited state or smaller mass of the scalar field  makes it easier for the emergence of translating point from the second-order transition to the first-order one. For the critical chemical potential $\mu_{c}$, we pointed out that the difference of $\mu_{c}$ between the consecutive states increases as the superfluid velocity $S_{x}/\mu$ increases. Interestingly, completely different from the holographic superfluid model with the ground state, the ``Cave of Winds" phase structure, which is associated with the emergence of a second superfluid phase, will disappear but the first-order phase transition occurs for the excited states, which indicates that the excited state will hinder the appearance of the Cave of Winds. Obviously, the excited state and superfluid velocity provide richer physics in the phase transition of the holographic model. This behavior is reminiscent of that seen for some condensed matter systems, where there exists a rich phase structure. For example, Deo, Peeters and Schweigert studied the mesoscopic superconducting disks with the ground state and metastable states in a perpendicular magnetic field, and observed the first or second order phase transitions for the normal to superconducting state, depending on the finite radius $R$ and thickness $d$ \cite{DeoPS2008}. Thus, our work may provide some useful information about the non-equilibrium superfluid (superconductor) which can be compared to the experiments in condensed matter physics. Moreover, we calculated the conductivity of the holographic superfluid system, and found that, similar to the excited states of holographic superconductors \cite{WangYQ2020}, there exist additional poles in Im[$\sigma(\omega)$] and delta functions in Re[$\sigma(\omega)$] arising at low temperature. On the other hand, we noted that increasing the number of nodes $n$ or superfluid velocity $S_{x}/\mu$ will decrease $\omega_{g}/T_{0}$, which tells us that the higher excited state or larger superfluid velocity results in the larger deviation from the expected relation in the gap frequency $\omega_{g}/T_{c}\approx8$. Although the present approach can extract the main physics of the holographic superfluid with the excited states and avoid the complex computation, it would be of great interest to extend the discussions beyond the probe limit. We will leave it for further study.

\begin{acknowledgments}

This work was supported by the National Key Research and Development Program of China (Grant No. 2020YFC2201400), National Natural Science Foundation of China (Grant Nos. 12275079, 12205060 and 12035005), and Postgraduate Scientific Research Innovation Project of Hunan Province (Grant No. CX20210472).

\end{acknowledgments}

\end{document}